\documentclass[%
 reprint,
superscriptaddress,
 amsmath,amssymb,
 aps,
prb,
]{revtex4-2}

\usepackage{graphicx}% Include figure files
\usepackage{dcolumn}% Align table columns on decimal point
\usepackage{bm}% bold math
\usepackage{verbatim}
\usepackage{enumerate}
\usepackage{epstopdf}
\usepackage{color}
\begin{document}

%\preprint{APS/123-QED}

\title{Floquet Engineering with Particle Swarm Optimization: Maximizing Topological Invariants}% Force line breaks with \\

\author{Shikun Zhang}
 \email{3120170459@bit.edu.cn}
\affiliation{School of Automation, Beijing Institute of Technology.}%Lines break automatically or can be forced with \\
\author{Jiangbin Gong}%
 \email{phygj@nus.edu.sg }
\affiliation{%
 Department of Physics, National University of Singapore.
}%


%\affiliation{
%\affiliation{
 %Third institution, the second for Charlie Author
%\affiliation{%


\date{\today}% It is always \today, today,

\begin{abstract}
It is of theoretical and experimental interest to engineer topological phases with very large topological invariants via periodic driving.  As advocated by this work, such Floquet engineering can be elegantly achieved by the particle swarm optimization (PSO) technique from the swarm intelligence family. With the recognition that conventional gradient-based optimization approaches are not suitable for directly optimizing topological invariants as integers, the highly effective PSO route yields new promises in the search for exotic topological phases, requiring limited physical resource. Our results are especially timely in view of two important insights from literature: low-frequency driving may be beneficial in creating large topological invariants, but an open-ended low-frequency driving often leads to drastic fluctuations in the obtained topological invariants. Indeed, using a simple continuously driven Harper model with three quasi-energy bands, we show that the Floquet-band Chern numbers can enjoy many-fold increase compared with that using a simple harmonic driving of the same period, without demanding more energy cost of the driving field. It is also found that the resulting Floquet insulator bands are still well-gapped, with the maximized topological invariants in agreement with physical observations from Thouless pumping.  The emergence of many edge modes under the open boundary condition is also consistent with the bulk-edge correspondence.   Our results are expected to be highly useful towards the optimization of many different types of topological invariants in Floquet topological matter.
%\begin{description}
%\end{description}
\end{abstract}

\maketitle

%\tableofcontents

\section{Introduction}
Topological matter \cite{RevModPhys.82.3045, RevModPhys.89.041004,Zhang2011} is associated with many potential applications and may bring the next revolution in material science to our daily life.  In particular, topological insulator materials accommodate robust conducting channels at their edges, and this feature can be crucial for lowering contact resistance in integrated circuits \cite{PhysRevLett.111.136801}, which is a fundamental goal in electronics. The same physics, when carried to other wave transport phenomena, holds promises in realizing a variety of topologically protected wave transport devices, such as topological acoustics \cite{Yang2015}, topological photonics \cite{Rech2013}, and  one-way waveguides without backward scattering \cite{PhysRevLett.113.113904}.  Because the number of topologically protected edge channels must agree with the bulk's topological invariants according to the principle of bulk-edge correspondence, it is of great theoretical and experimental interest to engineer topological phases with large topological invariants and hence scale up the number of topologically protected edge transport channels.

It has now been recognized that periodic driving can be a fruitful means to the generation of novel topological phases that are otherwise absent in static systems. The term ``Floquet topological matter" reflects this topical research line \cite{PRB2009,NP2011,Jiang2011,PhysRevLett.109.010601,Tong2013,Lindner2013,PlateroPRL,Ho2014,Rech2013,NJPRL}. Among many remarkable results, one particularly important observation is that periodic driving applied to simple model systems in low-frequency cases is more likely to create topological phases with large topological invariants \cite{Tong2013,Ho2014,zhou2014aspects}. As such, Floquet topological phases can potentially push the promises of topological matter to new limits. For example, via periodic driving and by exhaustive search in the low-frequency domain, we may now generate (i) many edge modes simultaneously for possible use of quantum information encoding \cite{Raditya2018,RadityaPRL2018}; (ii) a large number of topological corner modes in second-order topological systems \cite{Raditya2019}, and (iii) in the context of Floquet quantum Hall insulator, many chiral edge modes for significant enhancement of edge conductance \cite{YapPRB}. In the case of periodic quenching \cite{Ho2014,PhysRevB.93.184306, Zhou2018,PhysRevA.97.063603,Raditya2019}, the low-frequency driving is obviously equivalent to fixing the quenching period, while increasing the energy scale of the concerned Hamiltonians between which the system is periodically quenched. Indeed, using certain simple models \cite{Zhou2018}, it is shown that in the absence of physical constraints (such as energy cost), an arbitrarily large number of chiral edge modes can be produced. Note however, such periodic quenching models generally require instantaneous changes in the field, a feature that necessarily requires a huge number of harmonic components to realize the periodic driving field.

The open question is hence the following: how to systematically maximize the topological invariants of Floquet topological phases by optimizing the driving field, without relying on very strong driving fields or too many high harmonic components of a periodic driving field? To further appreciate the importance of this question, we further note that in the low-frequency limit, Floquet topological matter under open-ended driving tends to possess universal and violent fluctuations in their topological invariants  \cite{PhysRevLett.121.036402}. Evidently then, it is hopeful to obtain very large topological invariants by applying periodic driving, but to actually materialize them in a controlled and experimentally friendly manner needs active and smart engineering of the driving field.

In this work, we advocate an optimization routine to maximize the topological invariants of Floquet topological phases with a continuous periodic driving field, under bounded energy cost and using a limited number of high harmonic components. Remarkably, because topological invariants are integers, conventional gradient-based optimization approaches are not suitable to directly optimize these integers (due to the lack of a gradient).  Instead, we advocate to use the particle swarm optimization (PSO) technique \cite{8561256} from the swarm intelligence family \cite{996017} to achieve our goal. Though we report our specific results using an explicit model, our methodologies are equally applicable in optimizing any type of topological invariants.

Our case study below is built on a previously studied model \cite{zhou2014aspects}, namely, the continuously driven harper model (CDHM). The CDHM represents a Floquet topological insulator model, depicting a one-dimensional topologically nontrivial lattice plus a synthetic (or a decoupled transverse) dimension.  Previously, CDHM is modulated by a single-frequency cosine field and it was indeed observed that as the driving frequency decreases, there are circumstances in which relatively large topological Chern numbers are generated.  Here we replace the simple open-ended driving field by a periodic driving field with the same basic period, costing the same or even less energy, and with a limited number of high harmonic components. We resort to PSO to optimize the relative amplitudes and phases of these high harmonics to maximize the obtained Floquet  band Chern numbers. As shown below, the results indicate that the PSO-based routine is highly effective, thus further advancing ongoing studies of Floquet topological phases with an optimal control perspective.

The paper is organized as follows. Section II illustrates our optimization scheme on a model-independent level. In Section III, we apply our schemae to the CDHM. Specifically, we report the maximized Floquet-band Chern number and the associated optimized driving fields. We also perform adiabatic pumping to confirm the physical relevance of the maximized Chern numbers. We shall examine the topological edge modes of the system under the open boundary condition. Interesting insights into the role of the driving field emerge as we look into the Floquet effective Hamiltonians. Section IV concludes this paper.

%In \cite{PhysRevB.93.184306}, where the motivation of producing larger topological invariant is also recognized, the authors propose a scheme of periodic quenching which yields Chern number $\vert\mathcal{C}\vert=7$ under certain parameter regimes. However, an optimization procedure is not explicitly involved in that article.

%We first illustrate our general PSO approach, {\it independent} of any physical models.  That is, we assume an arbitrary Hamiltonian subject to periodic driving.  We also assume that a particular topological integer is identified as the target to maximize.  Our optimization goal is then to retrieve an engineered periodic driving field to produce large, and hopefully optimal, topological invariants for the resulting Floquet topological phases.  For the sake of simplicity, we assume that only one driving field is present (which is hence a periodic scalar function), though our approach is readily applicable to cases with multiple driving fields.

\section{The Optimization Scheme}

\subsection{The optimization problem}
A general formulation of the optimization problem will be presented in this subsection. Given a driving period $T$, the periodic field is denoted as $u_T(t)=u_T(t+T)$. Periodicity implies that $u_T(t)$ can be expanded in terms of Fourier components. In this work, we restrict our discussion to driving fields expandable by a finite number of Fourier components:
\begin{equation}
    u_T(t)= \frac{a_0}{2}+\sum_{n=1}^{N} a_n \cos(\frac{2n\pi }{T}t)+b_n \sin(\frac{2n\pi }{T}t),
\end{equation}
where $N$ is the highest order under consideration. The Fourier coefficients $(a_0, a_1,..., a_N, b_1,...,b_N)$ will be concisely denoted as $\mathbf{x}\in R^{2N+1}$. The energy of a field is defined as $\Vert{u_T}\Vert=\int_0^T u_T^2(t)dt$.  Through straightforward calculations, the Fourier coefficients of a field with energy cost less or equal to $E$ should satisfy:
\begin{equation}
    a_0^2/2+\sum_{n=1}^N a_n^2+b_n^2 \leq E.
\end{equation}
This inequality defines a region $\mathcal{D}(E) \subset R^{2N+1}$, which will serve as the constraint of optimization.

It is clear from the above discussion that when $T$ is fixed, Fourier coefficient vector $\mathbf{x}$ completely characterizes driving field $u_T$, which in turn pins down the time-dependent Hamiltonian $H(u_T(t))$ and one period Floquet operator $U(T,0)=\hat{C}e^{-i\int_0^TH(u_T(t))dt}$, where $\hat{C}$ stands for time ordering. Starting from $U(T,0)$, the calculation of topological invariants takes different paths depending on their definitions and models under study, but they are still completely determined by $\mathbf{x}$.

Floquet systems generally admit different quasienergy bands, each having its own topological invariant. Since magnitude is the primary concern of this work, we can adopt a scalar measure by summing up absolute values or squares of the topological invariants on each Floquet band (or whatever transformation that yields positive numbers). Once this measure is fixed, we will have defined a function $F(\mathbf{x};H(u_T))$ mapping Fourier coefficients to a scalar measuring how large the topological invariants are. It should be noted that $F(\mathbf{x};H(u_T))$ is dependent on $H(u_T)$, the driving-field modulated Hamiltonian.

We are now in the position to state the optimization problem: find $\mathbf{x} \in \mathcal{D}(E)$, such that $F(\mathbf{x};H(u_T))$ is as large as possible.

\subsection{PSO for maximizing topological invariants}
In general there is no analytical solution to the optimization problem stated above.  As a matter of fact, the objective function may be so intricately correlated with its input variables that it can almost be viewed as a black box. Furthermore, due to the integer nature of topological invariants (i.e. Chern numbers, winding numbers), and the fact that small adjustments of the field may not lead to topological phase transitions, the value of $F(\mathbf{x};H(u_T))$ forms patches of discrete "platforms". This feature significantly hinders the application of numerical optimization techniques such as Gradient Descent (and possibly all optimization algorithms where gradient information is needed), insofar as on flat plateaus the gradient is zero everywhere and these algorithms will get stuck at the first step.

\begin{figure}
\includegraphics{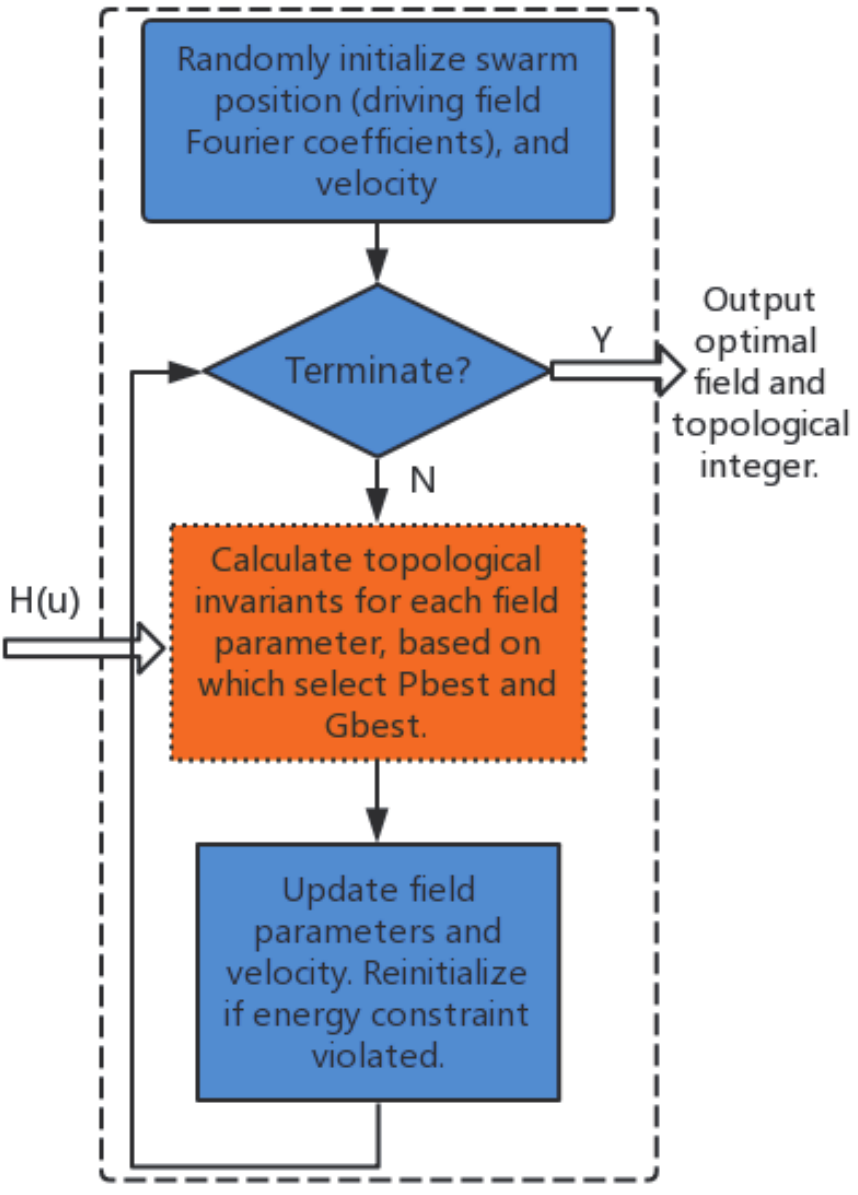}
\caption{Schematic of our PSO optimization approach to the generation of large topological invariants by engineering a periodic driving field. Our approach can be applied to maximize any type of topological invariants. }
\label{fig:epsart}
\end{figure}

However, PSO, which belongs to the category of evolutionary/swarm-intelligence optimization techniques, is non-gradient based and needs nothing more than the value of $F(\mathbf{x};H(u_T))$. In other words, even if we are encountered with a black box optimization problem, as long as the value of objective function can be computed, PSO can be implemented.  The optimization procedure goes as follows:

\begin{enumerate}
    \item Randomly initialize the positions of $M$ vectors (particles): $\mathbf{x}_1,\mathbf{x}_2,...,\mathbf{x}_M$ in $\mathcal{D}(E)$. Randomly set their initial velocities $\mathbf{v}_1^0,\mathbf{v}_2^0,...,\mathbf{v}_M^0$. For each $j\in \{1,...,M\}$, let $\text{Pbest}_j=\mathbf{x}_j$.

    \item For each $\mathbf{x}_j$, $j\in \{1,...,M\}$, compute $F(\mathbf{x}_j;H(u_T))$. If $F(\mathbf{x}_j;H(u_T))>F(\text{Pbest}_j;H(u_T))$, then let $\text{Pbest}_j=\mathbf{x}_j$. Denote $j^*=\arg \max _{j} F(\text{Pbest}_j;H(u_T))$, and select $\text{Gbest}=\text{Pbest}_{j^*}$.

    \item Update the position and velocity of each particle by the following iteration law:
\begin{align}
\qquad&\mathbf{v}_j\!=\!w\mathbf{v}_j\!+\!c_1 r_1(\text{Pbest}_jk\!-\!\mathbf{x}_j)\!+\!c_2 r_2(\text{Gbest}\!-\!\mathbf{x}_j) \\ \nonumber
\qquad&\mathbf{x}_j\!=\!\mathbf{x}_j+\mathbf{v}_j
\end{align}

\item For each $j\in \{1,...,M\}$, if $\mathbf{x}_j$ is outside $\mathcal{D}(E)$, randomly reinitialize it inside $\mathcal{D}(E)$.

\item If the termination criteria is met, output Gbest and end the procedure. Otherwise, go to step No. 2.

\end{enumerate}

In step No. 3, $w$, $c_1$, and $c_2$ are parameters to be fixed in advance. $r_1$ and $r_2$ obey uniform distribution in $[0,1]$ and are sampled in each iteration. Clearly, $\text{Pbest}_j$ stands for the position where particle $j$ attained the largest objective value in its own history, and $\text{Gbest}$ represents the largest objective value position found by all particles. Intuitively, this optimization procedure mimics the forage behavior of birds, where each bird decides its next move by retaining some of its current velocity (tuned by $w$), flying towards its own best position (tuned by $c_1$), and towards the best position of the flock (tuned by $c_2$). Such collective behavior somehow endows the swarm with ``intelligence" which is not manifested on an individual level. For example, when a particle falls into a local minima or maxima, it will be trapped there should it be on its own. However, with the presence of a swarm, this particle is able to move out from the trap following the interation law in step No. 3.

Thanks to swarm cooperation and the stochastic nature of this algorithm, the particles are able to hop between platforms searching for larger topological invariants (not aimlessly because the imformation of best positions is consumed). This may hopefully lead to global optimization.

\begin{table*}
\caption{\label{tab:table3}Floquet band Chern numbers in our three-band CDHM  model under optimized fields found from PSO, with given values of $J,V$ and driving period of $T=2$. Results here should be compared with parallel results in Table II using the simple cosine driving field. It is seen that as the values of $J,V$ increase, PSO becomes highly effective with many-fold increase in the Chern numbers.   }
\begin{ruledtabular}
\begin{tabular}{ccccccccccc}
 %&\multicolumn{2}{c}{$D_{4h}^1$}&\multicolumn{2}{c}{$D_{4h}^5$}\\
 $J=V$ & 0.5 & 1.5 & 2.5 & 3.5 & 4.5 & 5.5 & 6.5 & 7.5 & 8.5 & 9.5\\
 \hline
 $C_1$ & -2 & -2 & -2 & -4 & -4 & -5 & -5 & -7 & -7 & -11 \\
 $C_2$ &  4 &  4 &  4 &  8 &  8 & 10 & 10 & 14 & 14 &  22 \\
 $C_3$ & -2 & -2 & -2 & -4 & -4 & -5 & -5 & -7 & -7 & -11 \\
\end{tabular}
\end{ruledtabular}
\end{table*}

\begin{table*}
\caption{\label{tab:table3} Floquet band Chern numbers in our three-band CDHM  model using a simple cosine field with given values of $J,V$ and driving period of $T=2$. Results here should be compared with parallel results in Table I. }
\begin{ruledtabular}
\begin{tabular}{ccccccccccc}
 %&\multicolumn{2}{c}{$D_{4h}^1$}&\multicolumn{2}{c}{$D_{4h}^5$}\\
 $J=V$ & 0.5 & 1.5 & 2.5 & 3.5 & 4.5 & 5.5 & 6.5 & 7.5 & 8.5 & 9.5\\
 \hline
 $C_1$ & -2 & -2 & -2 & -4 & -4 & -2 &  4 &  4 & -2 & -2 \\
 $C_2$ &  4 &  4 &  4 &  8 &  8 &  4 & -8 & -8 &  4 &  4 \\
 $C_3$ & -2 & -2 & -2 & -4 & -4 & -2 &  4 &  4 & -2 & -2 \\
\end{tabular}
\end{ruledtabular}
\end{table*}

\begin{table*}
\caption{\label{tab:table3}Floquet band Chern numbers in our three-band CDHM under optimized fields found from PSO, with different choices of $T$ for fixed $J=3$, $V=4$. Results here should be compared with parallel results in Table IV using the simple cosine driving field.
 It is seen that as $T$ increases, PSO becomes highly effective with many-fold increase in the Chern numbers.}
\begin{ruledtabular}
\begin{tabular}{cccccccccccccc}
 %&\multicolumn{2}{c}{$D_{4h}^1$}&\multicolumn{2}{c}{$D_{4h}^5$}\\
 $T$ & 4.5 & 4.8 & 5.1 & 5.4 & 5.7 & 6 & 6.3 & 6.6 & 6.9 & 7.2 & 7.5 & 7.8 & 8.1\\
 \hline
 $C_1$ & -8 & -7 & -10 & -10 & -10 & -14 & -11 & -10 & -14 & -8 & -20 & -11 & -11\\
 $C_2$ & 16 & 14 &  20 &  20 &  20 &  28 &  22 &  20 &  28 & 16 &  40 &  22 &  22\\
 $C_3$ & -8 & -7 & -10 & -10 & -10 & -14 & -11 & -10 & -14 & -8 & -20 & -11 & -11\\
\end{tabular}
\end{ruledtabular}
\end{table*}

\begin{table*}
\caption{\label{tab:table3}Floquet band Chern numbers in our three-band CDHM using a simple cosine field with different choices of $T$ for fixed $J=3$, $V=4$. Results here should be compared with parallel results in Table III.  }
\begin{ruledtabular}
\begin{tabular}{cccccccccccccc}
 %&\multicolumn{2}{c}{$D_{4h}^1$}&\multicolumn{2}{c}{$D_{4h}^5$}\\
 $T$ & 4.5 & 4.8 & 5.1 & 5.4 & 5.7 & 6 & 6.3 & 6.6 & 6.9 & 7.2 & 7.5 & 7.8 & 8.1\\
 \hline
 $C_1$ & 4 & -2 & -2 & -2 & -2 & 4 & 4 & -8 & -2 & 4 & 4 & 4 & -2\\
 $C_2$ & -8 & 4 &  4 &  4 &  4 & -8 & -8 & 16 &  4 & -8 & -8 &-8 & 4\\
 $C_3$ & 4 & --2 & -2 & -2 & -2 & 4 & 4 & -8 & -2 & 4 & 4 & 4 & -2\\
\end{tabular}
\end{ruledtabular}
\end{table*}

In Fig.~1, we illustrate the idea proposed at the beginning of this section, where our scheme is visualized as a white box. The PSO inside the box can start to function  once the objective function is computed, which is the case if the model Hamiltonian is given as input. Then, the output Fourier coefficients determine the driving field we seek to optimize.

\section{Application}

In this section, we apply the optimization scheme outlined in Section II to CDHM \cite{zhou2014aspects}. The result of optimization is reported afterwards.

\subsection{The CDHM}
The CDHM considered in \cite{zhou2014aspects} is depicted by the following Hamiltonian:
\begin{equation}
    H(t)=J\cos(\hat{k})+V\cos(2\pi\alpha\hat{m}-\beta)\cos(\Omega t),
\end{equation}
where $\hat{k}$ denotes quasimomentum operator whose eigenvalues belong to [0,$2\pi$), and $\hat{m}$ represents position operator. Meanwhile, $\alpha=p/q$ is a rational number and $\beta$ is a phase shift parameter. In this work, we choose $p=1, q=3$ to yield a 3-band Floquet insulator model, with the $\beta$ being the synthetic dimension. With $\beta$ and the conserved quasimomentum $k$, the Chern numbers of the each of three Floquet bands can be obtained.

The previous version of CDHM adopts a simple cosine form for the driving field, with the single frequency $\Omega$. To maximize the topological Chern numbers we replace this cosine field with an engineered periodic field (with equal or less energy). With the field (periodic in $T$) to be optimized denoted as $u_T(t)$, the model under consideration becomes:
\begin{equation}
    H(u_T(t))=J\cos(\hat{k})+V\cos(2\pi\alpha\hat{m}-\beta)u_T(t).
\end{equation}
Its Floquet operator reads:
\begin{equation}
    U_{\text{CDHM}}(T,0)=\hat{C}e^{-i\int_0^TH(u_T(t))dt},
\end{equation}
where $\hat{C}$ stands for time ordering. Under the periodic boundary condition (PBC), this Floquet operator commutes with the translational operator $\hat{T}_k=e^{-iq\hat{k}}$. Therefore, for each fixed $\beta$, these two operators share a common set of eigenvectors. The Floquet eigen-equations now reads:
\begin{eqnarray}
U_{\text{CDHM}}(T,0)|\psi\rangle=e^{i\omega}|\psi\rangle,
\end{eqnarray}
with
\begin{eqnarray}
T_k|\psi\rangle=e^{i\phi}|\psi\rangle,
\end{eqnarray}
where $\omega$ is apparently the Floquet eigenphase and $\phi$ is simply given by $-qk$. A Brillouin zone (BZ) can be defined by scanning both the conserved quasi-momentum $\hat{k}$ and the phase shift parameter $\beta$ from $0$ to $2\pi$, based on which the Floquet operator $U_{\text{CDHM}}(T,0)$ yields $q=3$ bands \cite{zhou2014aspects}. Let us denote the eigenvector of the $n\text{th}$ band by $\left|\psi_{n}(\phi, \beta)\right\rangle$. Then, the Chern number of band $n$ is given by:
\begin{equation}
    C_{n}=\frac{i}{2 \pi} \oint d \mathbf{k} \cdot\left\langle\psi_{n}(\phi, \beta)\left|\nabla_{\mathbf{k}}\right| \psi_{n}(\phi, \beta)\right\rangle \quad \mathbf{k}=(\phi, \beta),
\end{equation}
which is the topological invariant under consideration for CDHM. A proper measure for the magnitude of Chern numbers of all bands is $\sum_n C_n^2$. Following the discussion in Section II, we should search for $\mathbf{x}$ (Fourier coefficients of the field) which makes $F(\mathbf{x};H(u_T))=\sum_n C_n^2$ as large as possible.

\begin{comment}
Clearly, Chern numbers are totally dependent on the details of the corresponding Floquet bands, which are in turn characterized by properties of Floquet operator $\hat{U}_{\text{CDHM}}(T,0)$. Since this operator is modulated by driving field $u(t)$, we are eventually able to engineer these Chern numbers by cherry picking the external field to be applied. In this article, periodic driving fields parametrized by their finite order Fourier expansion coefficients are considered. Let $T$ be the period of the driving field, then the expansion is expressed as:
\begin{equation}
    u(t)= \frac{a_0}{2}+\sum_{n=1}^{N} a_n cos(\frac{2n\pi}{T})+b_n sin(\frac{2n\pi}{T}),
\end{equation}
where $N$ is the highest order under consideration. We define the energy of the field as $\Vert{u}\Vert=\int_0^T u^2(t)dt$. Then, to have equal or less energy than the original cosine field, the coefficients should satisfy:
\begin{equation}
    a_0^2/2+\sum_{n=1}^N a_n^2+b_n^2 \leq 1.
\end{equation}
\end{comment}

\subsection{Optimization Results}
\begin{figure}
\includegraphics{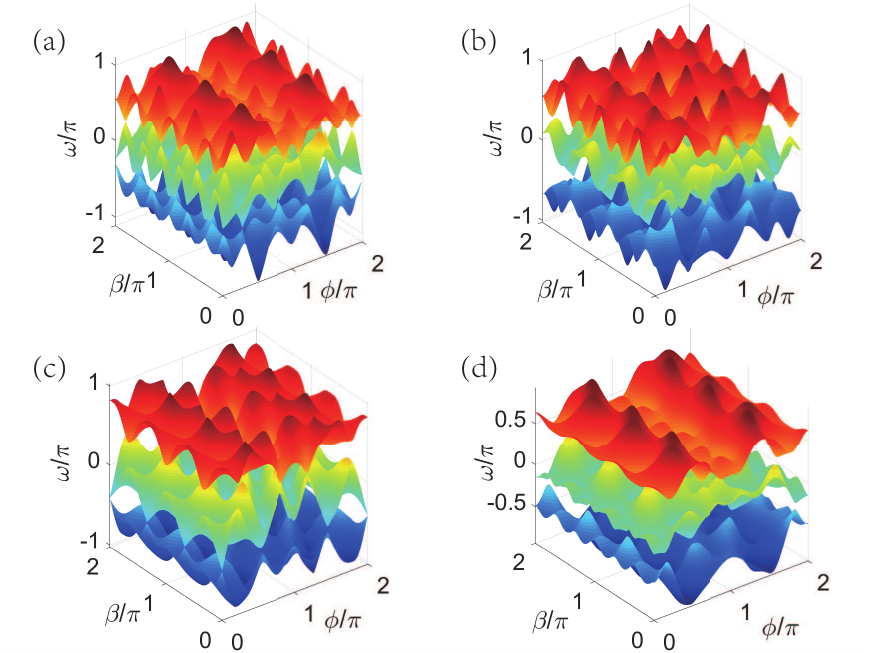}% Here is how to import EPS art
\caption{\label{fig:epsart} Quasi-energy bands for (a) $J=3$, $V=4$, $T=7.5$, with Chern numbers -20, 40, -20 (from bottom to top); (b) $J=3$, $V=4$, $T=6.9$, with Chern numbers -14, 28, -14; (c) $J=9.5$, $V=9.5$, $T=2$, with Chern numbers -11, 22, -11; (d) $J=8.5$, $V=8.5$, $T=2$, with Chern numbers -7, 14, -7;}
\end{figure}

\begin{figure}
\includegraphics{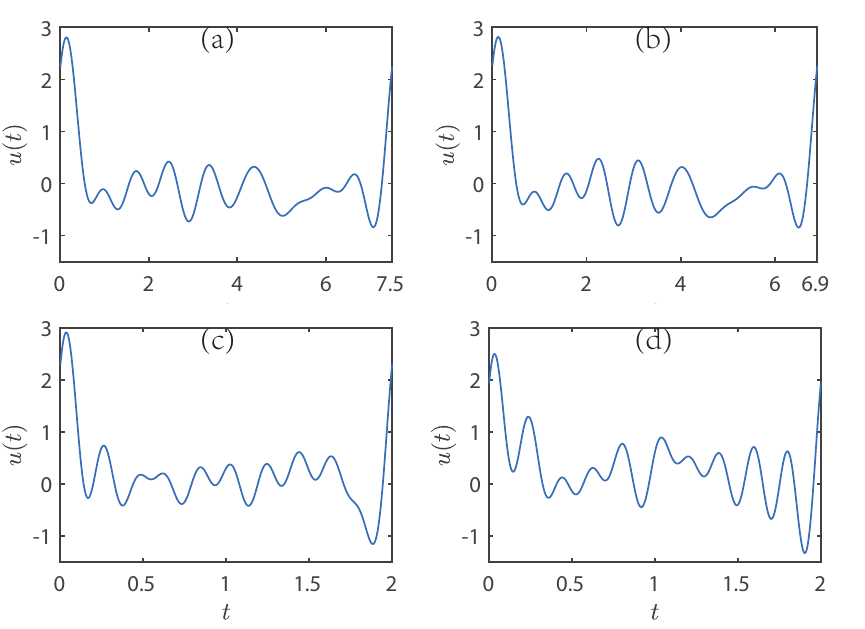}% Here is how to import EPS art
\caption{\label{fig:epsart} Driving field profiles found by PSO for (a) $J=3$, $V=4$, $T=7.5$; (b) $J=3$, $V=4$, $T=6.9$ (c) $J=9.5$, $V=9.5$, $T=2$; (d) $J=8.5$, $V=8.5$, $T=2$.}
\end{figure}

\begin{figure}
\includegraphics{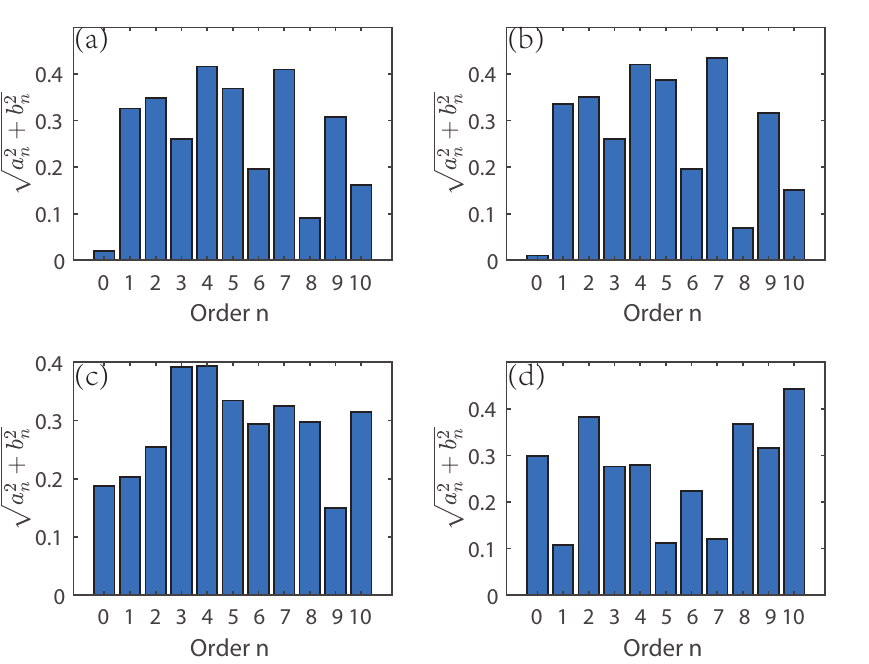}% Here is how to import EPS art
\caption{\label{fig:epsart} Magnitude of Fourier components of the driving fields found by PSO for (a) $J=3$, $V=4$, $T=7.5$; (b) $J=3$, $V=4$, $T=6.9$ (c) $J=9.5$, $V=9.5$, $T=2$; (d) $J=8.5$, $V=8.5$, $T=2$.}
\end{figure}

\begin{figure}
\includegraphics{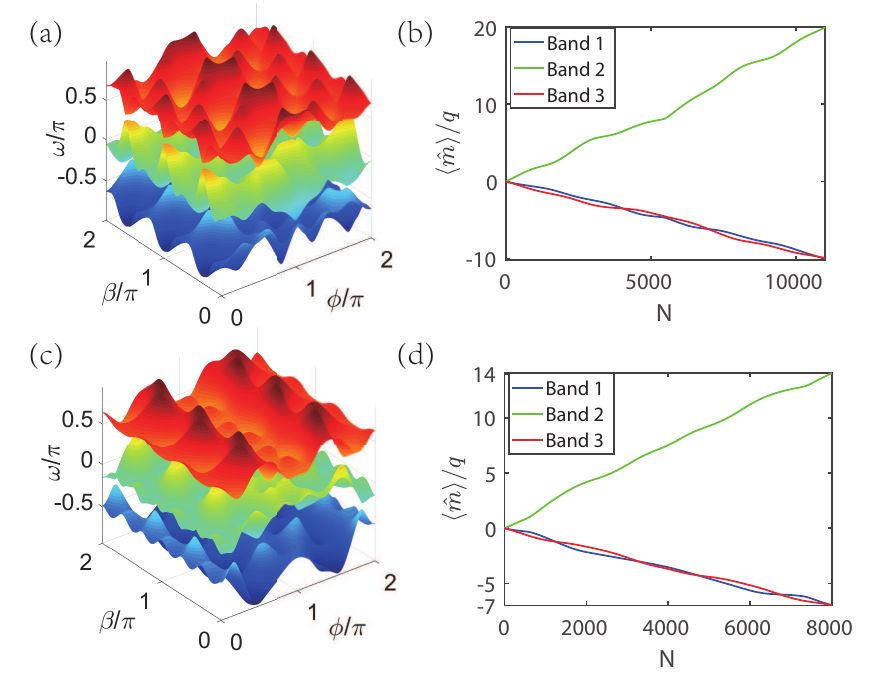}% Here is how to import EPS art
\caption{\label{fig:epsart} Adiabatic pumping for two sets of parameters. (a) Floquet spectrum for $J=3$, $V=4$, $T=5.4$. Chern numbers are: -10, 20, -10 (from bottom to top). (b) With system parameters $J=3$, $V=4$, $T=5.4$, the adiabatic cycle comprises of 11000 periods. Band numbers are ordered from bottom to top bands in (a). (c) Floquet spectrum for $J=8.5$, $V=8.5$, $T=2$. Chern numbers are: -7, 14, -7 (from bottom to top). (d) With system parameters $J=8.5$, $V=8.5$, $T=2$, the adiabatic cycle comprises of 8000 periods. Band numbers are ordered from bottom to top bands in (c). In panels (b) and (d), the expectation value of Wannier state center divided by $q$: $\langle \hat{m}\rangle/q $, is plotted against the number of periods. It is clear that by the end of both adiabatic cycles, the change in the wavepacket center approximately equals the Chern numbers of the associated Floquet bands.}
\end{figure}

The output of optimization, centered around the comparatively (with \cite{zhou2014aspects}) large Chern numbers found with respect to 23 sets of parameters for 3-band case, is reported in this subsection. %On certain sets of parameters where topological invariants are very large, the corresponding shapes of driving fields, the quasienergy bands and chiral edge modes under Open Boundary Conditions (OBC) which demonstrate bulk-edge correspondence are presented.

We have chosen $w=1.47$ and $c_1=c_2=1$ as parameters for PSO, and the highest order $N$ of Fourier expansion under consideration is 10. The energy of the field is restrained within $E=1$, which is equal to that of cosine field. The maximum number of iterations is set to be 120. However, the algorithm is terminated if the output of global best stays the same for 5 consecutive iterations.

\subsubsection{Optimized Chern numbers and New Topological phases}
Tables I-IV present maximized Floquet-band Chern numbers found by optimizing the driving fields via PSO, as compared with that obtained by simply applying the cosine field as in \cite{zhou2014aspects}. The 23 sets of parameters involve different combinations of $J$, $V$ in model (4), (5) and the period $T$ of driving field (1).  $C_1$, $C_2$, and $C_3$ represent Chern numbers of the 3 bands respectively.

In Tables I and II, we fix $T=2$ while keeping $J$ and $V$ the same, scanning them from 0.5 to 9.5. It is observed that significant improvements emerge when $J$ and $V$ are greater than 5.5, with $J=V=9.5$ giving the largest Chern numbers (in these two tables): -11, 22, -11.

More significant improvements are seen in Table III and IV, for fixed $J=3$ and $V=4$, whereas the period $T$ is scanned from 4.5 to 8.1. The effect of optimization is clearly demonstrated by larger topological invariants yielded by the optimized fields, which are collected in Table III, compared with the parallel results obtained  by using the open-ended cosine field in Table IV. It can be seen in Table III that Chern numbers as large as -20, 40, -20 ($J=3$, $V=4$, $T=7.5$) are found after optimization, which represents a 5-fold increase in comparison with the case with the same parameters under the cosine field in Table IV. Meanwhile, cases where $T$ equals 5.1, 5.4, 5.7, 6.9 and 8.1 also show 5 to 7-fold increase.  That is, as the driving period increases or as the field strength increases, the optimization outcome becomes much more pronounced than in cases with small $T$ or small field strength.

It is also noteworthy that by optimizing the driving field, we have obtained new topological phases that do not appear in Ref.~\cite{zhou2014aspects} using the simple cosine field. Indeed, Table II shows that when $J=V=6.5$ and $J=V=7.5$, the resulting Chern numbers are the same. This is because originally in Ref.~\cite{zhou2014aspects}, these two sets of parameters are shown to belong to the same topological phase. On the contrary, different Chern numbers can be obtained after optimization for these two cases (see Table I), clearly demonstrating the emergence of new topological states. Similarly, the creation of new topological phases are also substantiated by the results regarding $J=V=8.5$ and $J=V=9.5$ (compare Table I and II); $T=6$ and $T=6.3$ (compare Table III and IV); $T=7.2$, $T=7.5$, and $T=7.8$ (compare Table III and IV).

\subsubsection{Feautures of Quasienergy Bands and optimized Driving Fields}
We select 4 sets of parameters on which the PSO yields large Chern numbers and significant improvements over the open-ended case using the cosine field. Fig.~2 shows the Floquet spectrum obtained after optimization for these parameters. It can be seen that these bands are heavily wrinkled, a feature that is consistent with their very high Chern numbers.

Meanwhile, the shape of the driving fields that yield the Floquet spectrum in Fig.~2 is plotted in Fig.~3. The magnitude of each Fourier component of the fields in Fig.~3 is plotted in Fig.~4. In Fig.~4, the $x$-axis stands for the order of Fourier coefficients, and the $y$-axis stands for the square root of the sum of cosine and sine coefficients of corresponding orders. These figures do not show discernable patterns of the optimized fields. The Fourier components are neither concentrated near low or high frequencies, adding to the difficulties of guessing and hand-picking the driving field to be applied.  In some cases, even the zero-frequency component appears to be important.   We have also investigated what happens if we remove one of the harmonic components of the optimized field: in some cases the Chern numbers do not change (hence still in the same topological phase) whereas in some other cases it is affected drastically.
For all these reasons,  the need for implementing an optimization scheme like PSO is more clearly justified.

\begin{figure}
\includegraphics{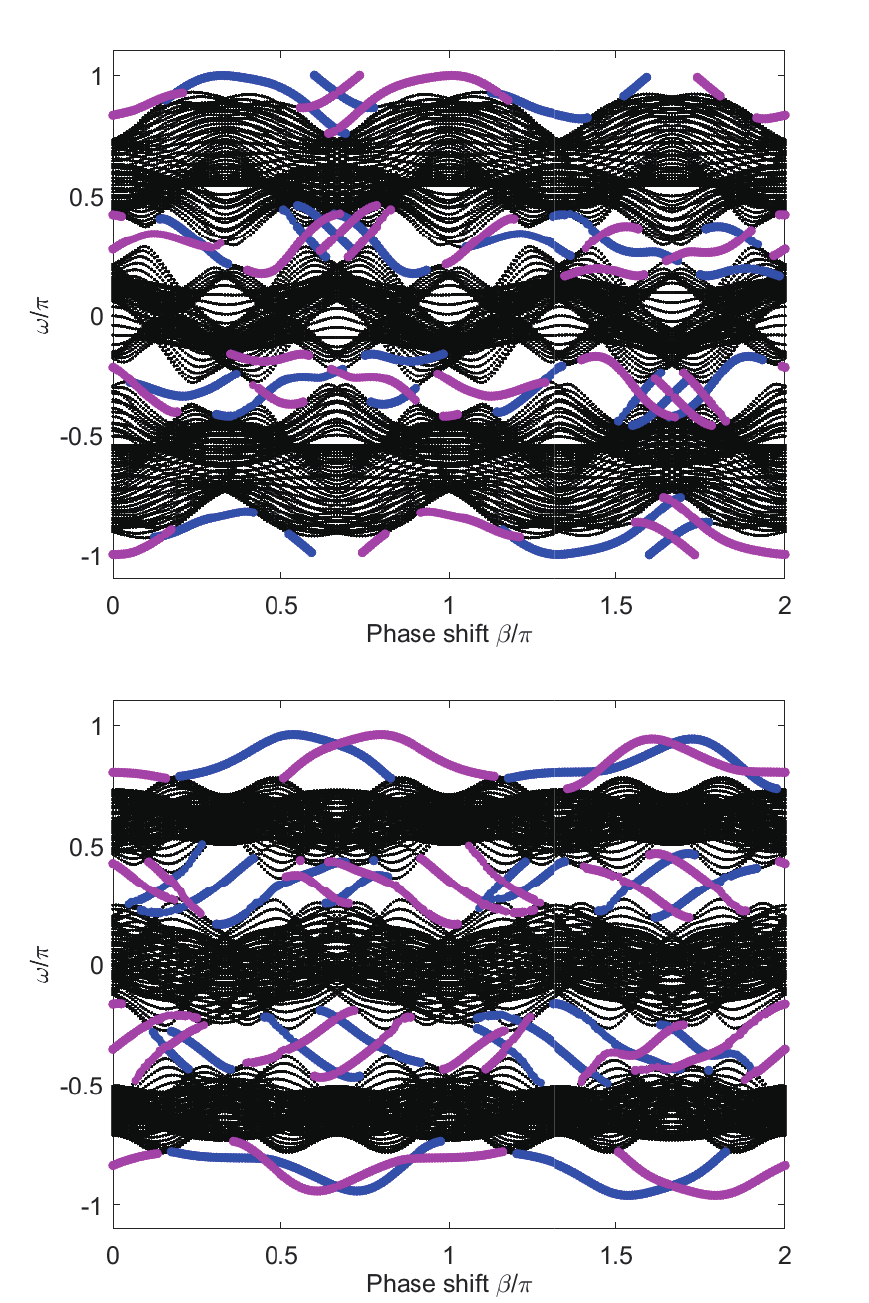}% Here is how to import EPS art
\caption{\label{fig:epsart} Floquet Eigenphase spectra versus phase shift $\beta$ under the open boundary condition for $J=8.5$, $V=8.5$, $T=2$ (top panel, Chern numbers -7, 14, -7); and $J=3$, $V=4$, $T=4.5$ (bottom panel, Chern numbers -8, 16, -8), each under the optimized driving fields found from PSO. Blue and margaret lines capture edge modes localized on the left and right, respectively. It can be verified that the difference between the net number of chiral edge mode pairs above and below a Floquet band equals its Chern number.}
\end{figure}

\subsubsection{Adiabatic Charge Pumping}
It was Thouless who developed the theory of quantized pumping for systems in the absence of periodic driving \cite{PhysRevB.27.6083}. By extending Thouless's theory, \cite{PhysRevLett.109.010601} revealed the implications of Floquet band topology on quantized adiabatic pumping. Under the same scenario in \cite{PhysRevLett.109.010601}, \cite{zhou2014aspects} numerically verified for the CDHM considered that a band Wannier state, which is a uniform superposition of all eigenstates on a topologically nontrivial Floquet band, would have its wavepacket center move over an integer number of lattice sites with $\beta$ adiabatically changing from 0 to $2\pi$. This integer is given by the Chern number of the associated band. Here we shall demonstrate computationally that our maximized Floquet band Chern numbers can be analogously connected with the physical process of Thouless pumping.

Let us denote the Wannier state of band $n$ at time $t$ by $W_n(t)$, then we expect to have:
\begin{equation}
    \langle W_n(NT)|\hat{m}|W_n(NT)\rangle-\langle W_n(0)|\hat{m}|W_n(0)\rangle = qC_n,
\end{equation}
where $T$ is the driving period, $N$ is the number of driving periods applied in an adiabatic cycle set to be very large, $C_n$ is the Chern number of band $n$, and $\hat{m}$ is the discretized position operator. Specifically, the Floquet band Wannier state is prepared following the method in \cite{PhysRevLett.109.010601}. Within each driving period, the phase shift $\beta$ is kept constant, but it is increased by a small amount in the following period, thus sweeping from $0$ to $2\pi$ in the complete adiabatic cycle. Fig.~5 exemplifies the quantized adiabatic pumping for our CDHM with optimized driving fields. We have adopted 40960 magnetic units cells to approximate the infinite length lattice in theory. It can be seen that the change of wavepacket centers is approximately equal to the Chern numbers of associated Floquet bands for large $N$, with accuracy gradually improving as more driving periods are used to complete one adiabatic cycle. The verification of this Thouless pumping indicates that, despite that the band Chern numbers are now increased by many times, the Floquet band gaps are still significantly large to yield quantized pumping.

\subsubsection{Edge Modes}
In this subsection, the Chern numbers for some selected examples, which are manifestations of bulk topology under PBC, will be related to the boundary modes under the open boundary condition. More specifically, we now consider our lattice to be of finite length $L$ with two open ends. The Hamiltonian reads:
\begin{align}
    H_{\text{OBC}}(t)=&\sum_{m=1}^{L-1} J(|m\rangle\langle m+1|+h . c .)\\ \nonumber
    +&\sum_{m=1}^{L} V \cos (2 \pi \alpha m-\beta) u_T(t)|m\rangle\langle m|,
\end{align}
where $|m\rangle$ denotes the position eigenstate at lattice $m$. The corresponding Floquet operator is expressed as:
\begin{equation}
    U_{\text{OBC}}(T,0)=\hat{C}e^{-i\int_0^TH_{\text{OBC}}(t)dt}.
\end{equation}
At each phase shift $\beta$, Floquet operator $U_{\text{OBC}}(T,0)$ is diagonalized. Scanning $\beta$ from $0$ to $2\pi$ then yields the eigenphase spectrum as a function of $\beta$. Fig.~6 depicts two such eigenphase spectra under the parameters and driving fields considered in the previous subsections. Open-end chains with length 150 are considered (unless stated otherwise). It is clear from Fig.6 that a large number of topological chiral edge modes indeed emerge between the gap of the bulk states, with some connecting the top and and the bottom Floquet bands. This confirms our initial goal: to maximize the topological invariants and hence generate a large number of topological edge modes. The abundance of chiral edge states made possible by an optimized driving field thus potentially facilitates a wide range of interesting applications based on edge state transport.

Moreover, the number of chiral edge modes is consistent with the Chern numbers of each Floquet band, demonstrating bulk-edge correspondence. Specifically, the net number of chiral edge mode pairs above a band substracts the number of the same quantity below it should equal the band's Chern number. For example, in the upper panel of Fig.~6, the Chern numbers of three Floquet bands are -7, 14, -7, from top to bottom. It can be checked that there are 7 pairs of edge modes connecting the middle and the top band. (The mode totally lying on the middle band colored in margaret does not count.) Meanwhile, there are also 7 pairs of edge modes with opposite chirality which traverse the middle and bottom bands. Counting each pair of top-middle edge mode as $+1$ and middle-bottom one as $-1$, the substraction $+7-(-7)=14$ yields the band's Chern number. When it comes to the top and bottom bands, we observe that they are actually connected by two pairs of edge modes with opposite chirality, giving a net number $0$ for the space above top and below bottom. The equations $0-(+7)=-7$ and $-7-0=-7$ thus agree with their Chern numbers respectively. Similarly, the lower panel of Fig.~6 also indicates bulk-edge correspondence.

\begin{figure}
\includegraphics{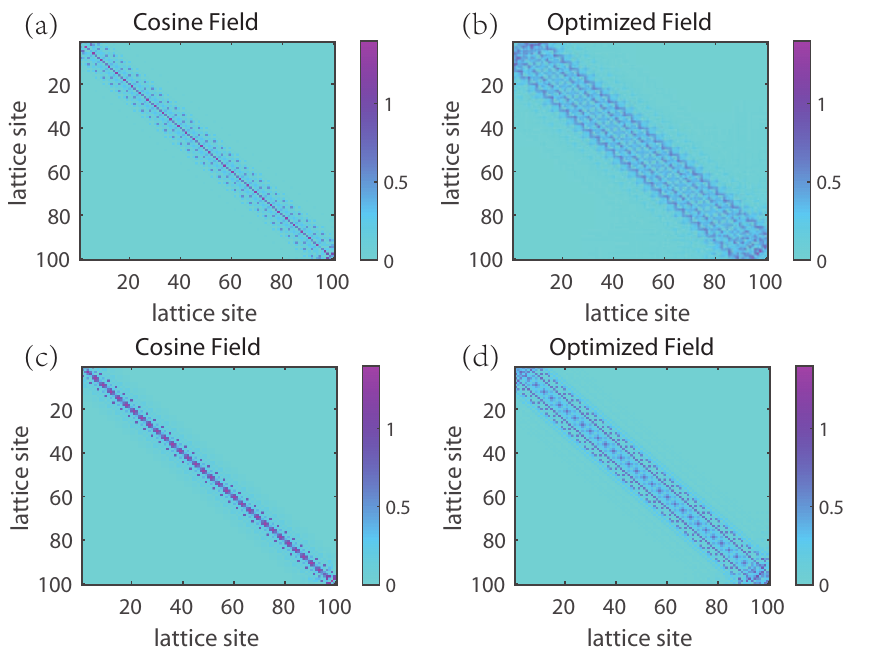}% Here is how to import EPS art
\caption{\label{fig:epsart} Floquet effective Hamiltonian matrix elements under cosine and optimized fields, presented in lattice space representation. The absolute value of each matrix element is calculated and plotted with different colors according to its magnitude. X and Y axes are lattice indices to label the matrix elements of the Floquet effective Hamiltonian. For (a) and (b), the system parameters are $J=3$, $V=4$ and $T=4.8$. For (c) and (d), they are $J=V=6.5$ and $T=2$. Enhanced middle-range hopping can be observed by comparing these panels.}
\end{figure}

\begin{figure}
\includegraphics{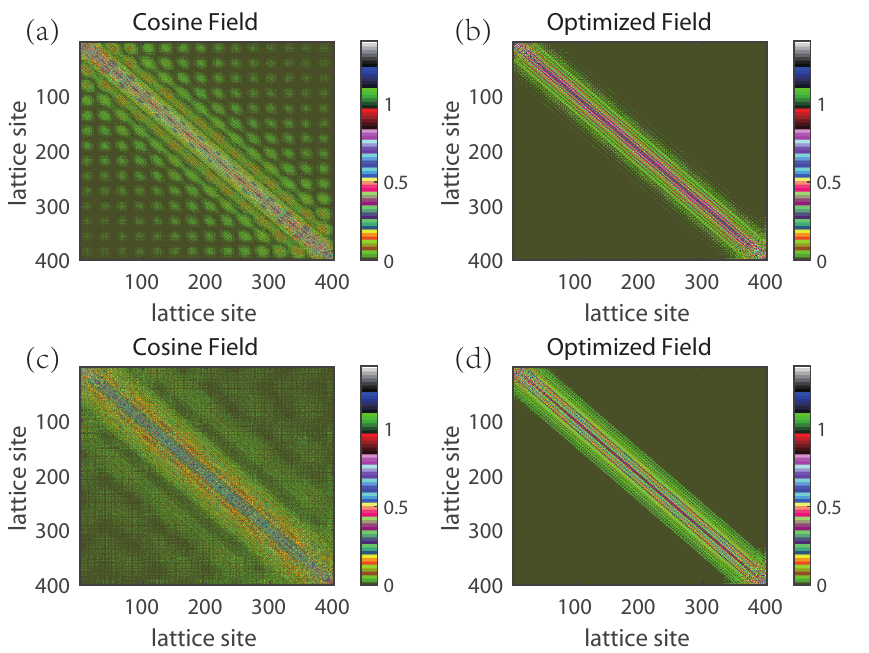}% Here is how to import EPS art
\caption{\label{fig:epsart} Floquet effective Hamiltonian matrix elements under cosine and optimized fields, presented in lattice space representation.  The absolute value of each matrix element is calculated and plotted with different colors according to its magnitude. X and Y axes are lattice indices to label the matrix elements of the Floquet effective Hamiltonian.  For (a) and (b), the system parameters are $J=3$, $V=4$ and $T=6.9$. For (c) and (d), they are $J=3$, $V=4$ and $T=8.1$. It is seen that very-long-range hopping is cleansed after optimization.}
\end{figure}

\begin{figure}[b]
\includegraphics{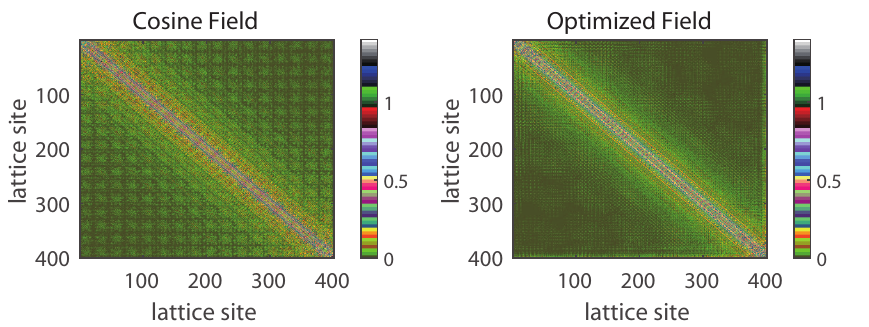}% Here is how to import EPS art
\caption{\label{fig:epsart} Floquet effective Hamiltonian matrix elements with system parameters $J=3$, $V=4$, $T=7.5$ for the simple cosine field (left) and the optimized field (right). It is seen that matrix elements indicating long-range hopping scatter among lattice sites far off the diagonal line in the case of the cosine driving field.  The case with the optimized field suppresses long-range hopping to some extent, yet the cleansing effect is not as thorough as in previous examples.}
\end{figure}

\subsubsection{Effective Hamiltonian}
With the driving period set to $T$, the effective Hamiltonian $H_{\text{eff}}^T$ is the average generator of Floquet operator $U_{\text{OBC}}(T,0)$, or formally:
\begin{equation}
  U_{\text{OBC}}(T,0)=e^{-iH_{\text{eff}}^T T}.
\end{equation}
Although $H_{\text{OBC}}(t)$ involves only nearest neighbor hoppings, $H_{\text{eff}}^T$ may effectively possess long-range hopping \cite{Tong2013}. This is one key insight highly relevant to topological nontriviality of Floquet bands. That is, $H_{\text{eff}}^T$, acting as an ``average" Hamiltonian within a period, potentially captures some long-range hopping, and hence its off-diagonal matrix elements in the lattice representation might decay to zero rapidly.  In order to digest how optimization of driving fields affects the hopping beyond nearest-neighbor lattices, we now examine the effective Hamiltonians under optimized fields and compare them with that under the simple cosine fields.

 Clearly, for static systems, the effective Hamiltonian is equal to the original Hamiltonian. If the original Hamiltonian only carries nearest neighbor hoppings, then all matrix elements of the effective Hamiltonian beyond one element away from the diagonal will vanish. By contrast, the effective Hamiltonian of Floquet systems is expected to show significantly nonzero elements within some distance from the diagonal line. Moreover, the magnitude distribution of these elements may strongly depend on the driving fields applied. We indeed computationally obtain the matrix elements of $H_{\text{eff}}^T$, for all parameter settings listed in Tables I-IV, with the results summarized below.

In this subsection, our numerical results are based on 400-site open boundary lattices (in contrast with 150 sites in the previous subsection) to fully observe the patterns of effective Hamiltonians. Firstly, compared with the case with a simple cosine driving field of the same period, our optimized fields which yield higher Chern numbers tend to enhance middle-range hopping, as manifested by the structure of the effective Hamiltonian matrix elements. For example, in Fig.~7 (a) where $J=3$, $V=4$, $T=4.8$ and with cosine field applied, it can be seen that the major hopping is constrained within a distance of 10 lattice sites. By contrast, in Fig.~7 (b) with the same $J$, $V$ and $T$ but under our optimized field, it is observed that significant hopping extends beyond a distance of 20 sites. A similar pattern is also obtained in Fig.~7 (c) and (d), where $J=V=6.5$ and $T=2$. That is, hopping around 10 lattice sites is enhanced after optimization. Since the pattern is already shown clearly when we look at the first 100 sites of the lattice, the rest are omitted to magnify the details we concern.

 Secondly, regarding very-long-range hopping, there are several parameter settings where the case with the cosine driving field produces effective Hamiltonians admitting non-vanishing elements far off the diagonal line. Our optimized cases with very large Chern numbers, on the contrary, present a ``cleansing" effect such that the magnitude of these far-off-diagonal elements are greatly suppressed. For instance, in Fig.~8 (a) ($J=3$, $V=4$, $T=6.9$) and (c) ($J=3$, $V=4$, $T=8.1$) where the cosine driving field is applied, slight but discernable hopping even extends to nearly 400 lattice sites, creating ``noisy" figures for the matrix elements of the effective Hamiltonians. In contrast, Fig.~8 (b) ($J=3$, $V=4$, $T=6.9$) and (d) ($J=3$, $V=4$, $T=8.1$) show that, after optimization, the areas far off the diagonal lines are much ``cleaner", suggesting the significant suppression of very-long-range hopping. In effect, our PSO routine has managed to let our driving field produce an effective Hamiltonian whose matrix elements focus heavily on 20-30 (an intermediate number) sites away from the diagonal line. (It is noted that, in this case, all 400 sites are plotted to present the bigger picture.)

We have also explore many other situations to further confirm the  interesting insight above.  Among many other cases, we find one single case where the above-mentioned cleansing effect is not that complete. In particular, for $J=3$, $V=4$ and $T=7.5$, it is mentioned previously that very large Chern numbers: -20, 40, -20 are obtained with PSO. Under the same parameter setting, the open-ended cosine driving field generates an effective Hamiltonian $H_{\text{eff}}^T$ that admit considerable long-range hopping for up to 400 sites away from the diagonal line. For our optimized field, some cleansing effects are still observed, but not so strong as other cases discussed above.  As shown from the structure of the matrix elements of the Floquet effective Hamiltonian in Fig.~9, certain very-long-range hopping, albeit with small amplitudes, is still present.  It remains to be understood why the cleansing effect here is not thorough in this particular case.
%This observation is illustrated by FIG.9.

\section{Conclusions}

In this work, we have advocated to use the PSO to efficiently generate Floquet topological phases with many-fold increase in the topological invariants as compared with open-ended simple harmonic driving of the same period, without demanding for more energy cost.   In presenting the explicit results, we have used a specific model. Nevertheless, it is evident that our PSO idea can be applied to maximize any type of topological integers in other systems, so long as the problem is not computationally too demanding.  To facilitate our calculations, we have deliberately chosen to examine cases with driving frequencies still not too low.  The observed trend is already clear: the PSO approach can be even more effective as we further approach the low-frequency regime.  The irregularity of optimized periodic driving fields in return supports the adoption of intelligent optimization techniques rather than any rational guess based on experience or physical intuition.

In terms of physics, we have also confirmed that the maximized topological invariants obey the bulk-edge correspondence and can be physically connected with Thouless pumping results. By analyzing the structure of the matrix elements of the Floquet effective Hamiltonians, we observe that the optimized driving field tends to induce an effective Hamiltonian with matrix elements heavily focused on intermediate-range hopping, whereas uncontrolled driving (including without driving) either has short-range hopping only, or has uncontrolled long-range hopping across too many lattice sites. We believe that such qualitative insights can be useful for developing more understandings of the optimization problem here.

\begin{acknowledgments}
J.G. acknowledges
fund support by the Singapore NRF grant No. NRF-NRFI2017- 04 (WBS No. R-144-000-378-281).
\end{acknowledgments}

\bibliography{PSO}% Produces the bibliography via BibTeX.

%\newpage

\end{document}